\documentclass[a4paper,twocolumn,pra,showpacs,aps]{revtex4-1}
\usepackage{epsfig,amsmath,amsfonts,amssymb}
\usepackage{amsmath,amsfonts,amssymb}
\usepackage{graphicx}
\usepackage{subfigure}
\usepackage{mathbbol}

\begin{document}

\title{Design of a Photonic Lattice Using Shortcuts to Adiabaticity}

\author{Dionisis Stefanatos}
\email{dionisis@post.harvard.edu}
%\altaffiliation[Current address:]{ 3 Omirou St., Sami, Kefalonia 28080, Greece.}
%\email{dionisis@post.harvard.edu}
\affiliation{3 Omirou St., Sami, Kefalonia 28080, Greece}

\date{\today}% It is always \today, today,

\begin{abstract}
In this article we use the method of shortcuts to adiabaticity to design a photonic lattice (array of waveguides) which can drive the input light to a controlled location at the output. The output position in the array is determined by functions of the propagation distance along the waveguides, which modulate the lattice characteristics (index of refraction, first and second neighbor couplings). The proposed coupler is expected to posses the robustness properties of the design method, coming from its adiabatic nature, and also to have a smaller footprint than purely adiabatic couplers. The present work provides a very interesting example where methods from quantum control can be exploited to design lattices with desired input-output properties.
\end{abstract}

%\pacs{05.45.Xt}% PACS, the Physics and Astronomy
\maketitle

%%%%%%%%%%%%%%%%%%%%%%%%%%%%%%%%%%%%%%%%%%%%%%%%%%%%%%%%%%%%%%%%%%%%%%%%%%%%%%%%
\section{INTRODUCTION}

\label{sec:intro}

Many recent studies have demonstrated that a family of photonic lattices, implemented in arrays of evenescently coupled waveguides \cite{Christ03}, provides the ideal testbed for a large variety of quantum phenomena \cite{Longhi09}. We mention the observation in these lattices of Bloch-like revivals \cite{Keil12,Corrielli13,Khazaei13}, Anderson localization of light \cite{Segev13,Giuseppe13,Lee14}, and the classical analog of coherent states and quantum correlations \cite{Keil11,Perez11}, to name a few. On the other hand, ideas from quantum mechanics have inspired the design of optical structures with desired properties. For example, a photonic lattice engineered to mimic the dynamics of a spin-chain has been used to achieve coherent quantum transport \cite{Perez13}, while the concept of supersymmetry has been exploited to implement mode-division multiplexing \cite{Miri13,Heinrich14}.

Recently, a class of such photonic lattices has been introduced, with dynamics analogous to a forced quantum harmonic oscillator \cite{Rodriguez14}. The mass, frequency and external force of the corresponding oscillator are determined by the index of refraction and the first and second neighbor couplings of the waveguides in the lattice, while all these parameters depend on the propagation distance along the waveguides. As pointed out in \cite{Rodriguez14}, such a quantum oscillator posses a Lewis-Riesefeld invariant \cite{Lewis69}, and the corresponding lattice exhibits the same symmetry. For quantum systems with this kind of invariants, a method called \emph{shortcuts to adiabaticity} has been proposed to engineer the output state \cite{Chen10,shortcuts13}. This method provides a path connecting the initial and final eigenstates of a parameter-dependent Hamiltonian (the parameter is usually the time, here is the propagation distance), without following the instantaneous eigenstates, producing thus an effectively adiabatic evolution. The method is in general quite robust \cite{Choi12} and has found applications in various physical contexts \cite{shortcuts13}. Some optics related applications, based on the principle of the adiabatic directional coupler \cite{Paspalakis06}, include the fast and robust mode conversion and splitting in multimode waveguides \cite{Tseng12,Chien13}, the efficient beam coupling in a three waveguide directional coupler \cite{Tseng13}, and the implementation of a compact and high conversion efficiency mode-sorting asymmetric Y-junction \cite{Martinez14}. In all these applications the device length is considerably reduced compared to the purely adiabatic schemes.

In the present article we use the method of shortcuts to adiabaticity to design a waveguide array of the type \cite{Rodriguez14} such that light entering the first waveguide is directed to a subset of waveguides at the output. The location of the output in the array can be controlled by functions of the propagation distance which modulate the lattice characteristics (index of refraction, first and second neighbor couplings). The proposed coupler is expected to posses the robustness properties of the design method \cite{Choi12,Tseng12,Chien13,Tseng13,Martinez14}, coming from its adiabatic nature \cite{Paspalakis06}, and also to require less waveguide length than purely adiabatic couplers \cite{Tseng12,Chien13,Tseng13,Martinez14}. We also observe that the suggested lattice model has another very interesting feature. If the modulating functions can be altered in time, it can also act as a switch, directing the output to different locations depending on the applied control. Although time-varying couplings cannot be implemented in the context of evanescently coupled waveguides because they depend on the geometry which is constant, we believe that this useful property can probably find application to other areas, since lattice systems are ubiquitous in physics and science in general. The present work provides also an interesting example where the design of a photonic lattice can be reduced to the design of a control system with inputs the lattice characteristics. This is a very promising approach since tools from computational optimal control can be exploited to design more sophisticated outputs, for example a perfect state transfer as described in \cite{Perez13}. We explain in more detail how this can be done in the conclusion, after we discuss the lattice equations.
We also note that from the perspective of networks science this work provides an example of how global, open-loop control (not feedback) can be used to drive a network coherently \cite{Stefanatos12,Li13} and not asymptotically in time \cite{Hoppensteadt99}. Finally, we point out that it is possible to classically emulate quantum coherent states by using the suggested photonic lattice.

In the next section we summarize the results of \cite{Rodriguez14}, where the mathematical model describing the specific class of photonic lattices is given. In section \ref{sec:design}, which is the main contribution of this paper, we show how shortcuts to adiabaticity can be used to design the output of such a lattice and we provide simulation results. In section \ref{sec:conclusion}, which concludes this article, we outline how the present work can be extended by using computational optimal control.

%%%%%%%%%%%%%%%%%%%%%%%%%%%%%%%%%%%%%%%%%%%%%%%%%%%%%%%%%%%%%%%%%%%%%%%%%%%%%%%%
\section{A PHOTONIC LATTICE ANALOGOUS TO THE HARMONIC OSCILLATOR}

\label{sec:model}

In this article we use the model of a semi-infinite photonic lattice introduced in \cite{Rodriguez14}. It is composed by a linear array of evanescently coupled waveguides numbered as $n=0,1,2,\ldots$, while the direction of light propagation is denoted by $z$. The complex field amplitude at the $n$th waveguide $C_n(z)$ satisfies the following differential equation
\begin{eqnarray}
\label{model}
i\partial_z C_n(z)+a_0(z)nC_n(z)&&\nonumber\\
+a_1(z)[f_{n+1}C_{n+1}(z)+f_{n}C_{n-1}(z)]&& \nonumber\\
+a_2(z)[g_{n+2}C_{n+2}(z)+g_{n}C_{n-2}(z)]&=&0
\end{eqnarray}
where $f_n=\sqrt{n}, g_n=\sqrt{n(n-1)}$ are functions of the position $n=0,1,2,\ldots$ of the waveguide in the array and $C_n(z)=0$ for $n<0$. The first term in the above equation corresponds to the change of $C_n(z)$ along the propagation direction while the second term corresponds to a linearly varying refractive index ($\sim n$) modulated by a $z$-dependent function $a_0(z)$. The third and fourth terms represent first and second neighbor couplings, respectively, and are modulated by the functions $a_1(z)$ and $a_2(z)$.

As pointed out in \cite{Rodriguez14}, if we define the state vector $|\phi(z)\rangle=\sum_{n=0}^\infty C_n(z)|n\rangle$, where the auxiliary states $|n\rangle$ are defined in the sequence, the system of differential equations (\ref{model}) is equivalent to the following Schr\"{o}dinger equation ($\hbar=1$)
\begin{equation}
i\partial_z |\phi(z)\rangle=\hat{\mathcal{H}}|\phi(z)\rangle,
\end{equation}
where the Hamiltonian $\hat{\mathcal{H}}$ is given by
\begin{equation}
\hat{\mathcal{H}}=-[a_0(z)\hat{a}^\dag \hat{a}+a_1(z)(\hat{a}+\hat{a}^\dag)+a_2(z)(\hat{a}^2+\hat{a}^{\dag 2})]
\end{equation}
in terms of the creation and annihilation operators satisfying
\begin{equation}
\label{ladder}
\hat{a}|n\rangle=\sqrt{n}|n-1\rangle,\qquad\hat{a}^\dag|n\rangle=\sqrt{n+1}|n+1\rangle.
\end{equation}
Expressing these operators in terms of the normalized position and momentum operators
\begin{equation}
\label{creation}
\hat{a}=\frac{\hat{q}+i\hat{p}}{\sqrt{2}},\qquad\hat{a}^\dag=\frac{\hat{q}-i\hat{p}}{\sqrt{2}},
\end{equation}
the above Hamiltonian becomes
\begin{equation}
\label{H_z}
\hat{\mathcal{H}}=-\left[\frac{\hat{p}^2}{2M(z)}+\frac{M(z)\Omega^2(z)\hat{q}^2}{2}-F(z)\hat{q}-\frac{a_0(z)}{2}\right],
\end{equation}
where the mass, frequency and driving force depend on the propagation distance $z$ and are related to the lattice parameters through the relations
\begin{eqnarray}
\label{M}M(z)&=&\frac{1}{a_0(z)-2a_2(z)},\\
\label{W2}\Omega^2(z)&=&a_0^2(z)-4a_2^2(z),\\
\label{F}F(z)&=&-\sqrt{2}a_1(z).
\end{eqnarray}
Note that because of the ladder relations (\ref{ladder}) and the specific choice of position and momentum operators (\ref{creation}), the auxiliary states $|n\rangle$ are the eigenstates of a normalized harmonic oscillator
\begin{equation}
\label{eigenstates}
|n\rangle=\phi_n(q)=\frac{1}{(2^nn!)^{1/2}\pi^{1/4}}e^{-q^2/2}H_n(q),
\end{equation}
where $H_n$ is the Hermite polynomial of degree $n$. For a general state $|\phi(z)\rangle$, the complex amplitude in the $n$th waveguide and at the position $z$ is
\begin{equation}
\label{amplitude}
C_n(z)=\langle\phi(z)|n\rangle.
\end{equation}

Observe that the last term in (\ref{H_z}) simply introduces a common phase factor. If we define the wavefunction $|\psi(-z)\rangle=e^{\frac{i}{2}\int_0^z{a_0(\zeta)d\zeta}}|\phi(z)\rangle$ and make the change of variable $t=-z$, then we easily find that $|\psi(t)\rangle$ satisfies
%\begin{equation}
%\label{Oscillator}
%i\partial_t |\psi(t)\rangle=\left[\frac{\hat{p}^2}{2m(t)}+\frac{m(t)\omega^2(t)\hat{q}^2}{2}-f(t)\hat{q}\right]|\psi(t)\rangle,
%\end{equation}
\begin{equation}
\label{evolution}
i\partial_t |\psi(t)\rangle=\hat{H}|\psi(t)\rangle,
\end{equation}
where
\begin{equation}
\label{Oscillator}
\hat{H}=\frac{\hat{p}^2}{2m(t)}+\frac{m(t)\omega^2(t)\hat{q}^2}{2}-f(t)\hat{q}
\end{equation}
and $m(t)=M(-t), \omega(t)=\Omega(-t), f(t)=F(-t)$. This is the equation of a quantum harmonic oscillator with time-dependent mass, frequency, and external driving force.

%%%%%%%%%%%%%%%%%%%%%%%%%%%%%%%%%%%%%%%%%%%%%%%%%%%%%%%%%%%%%%%%%%%%%%%%%%%%%%%%
\section{ENGINEERING THE LATTICE OUTPUT USING SHORTCUTS TO ADIABATICITY}

\label{sec:design}

\begin{figure}[t]
\centering
\includegraphics[width=0.9\linewidth]{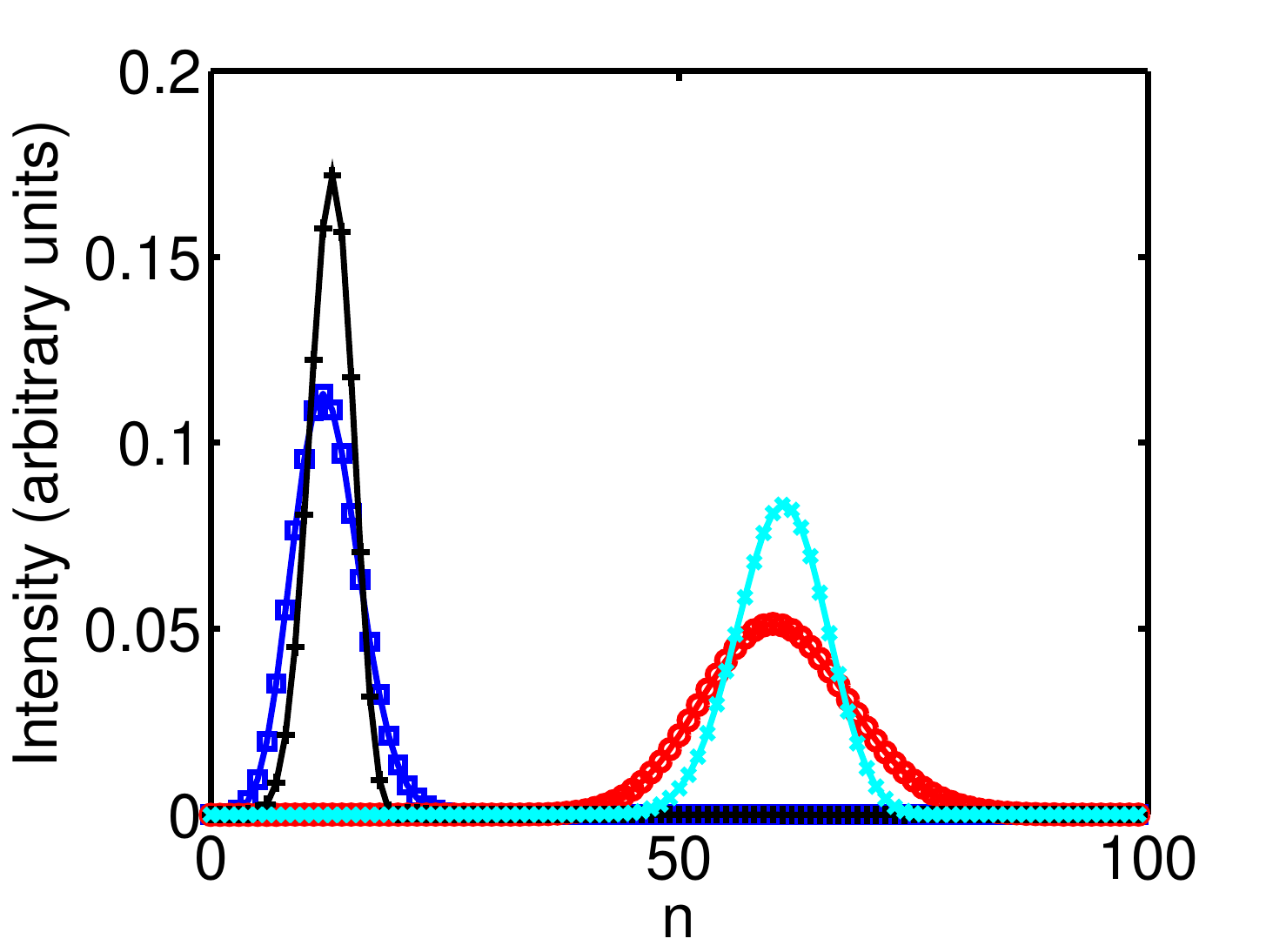}
\caption{(Color online) Output intensities in a photonic lattice composed by 100 waveguides, for light injected in the $0$th waveguide and propagated for a distance $Z=10$. Each curve corresponds to a different choice of the lattice parameters, given in Figs. \ref{fig:transport5}, \ref{fig:transport6}. The blue-squares curve corresponds to the target final state (\ref{states}) with $d=5$, while the red-circles curve to $d=11$. The black-crosses curve corresponds to the target final state (\ref{enhanced_final}) with $d=5,\sigma=0.6$, while the cyan x-marks curve to $d=11,\sigma=0.6$}.
\label{fig:outputs}
\end{figure}

The design of photonic lattices of the class described in the previous section is reduced to the appropriate choice of the modulating functions $a_i(z), i=1,2,3$, such that the input state at $z=0$ is mapped to the desired output at the final distance $z=Z$. In this paper we consider input only in the $0$th waveguide, so $C_0(0)=1$ and $C_n(0)=0$ for $n>0$, while the desired output is a displaced version of the input. The corresponding initial and final states are
\begin{equation}
\label{states}
|\psi(0)\rangle=\frac{1}{\pi^{1/4}}e^{-q^2/2},\qquad|\psi(-Z)\rangle=\frac{1}{\pi^{1/4}}e^{-(q+d)^2/2},
\end{equation}
where $d$ is the displacement parameter. Note that the final state is a coherent state with eigenvalue of the annihilation operator $\alpha=-d/\sqrt{2}$. Using the generating function of the Hermite polynomials \cite{Merzbacher98}
\begin{equation}
\label{generating}
e^{-x^2+2xq}=\sum_{n=0}^\infty\frac{H_n(q)}{n!}x^n
\end{equation}
with $x=-d/2$ we can easily express the final state in terms of the states $|n\rangle$ given in Eq. (\ref{eigenstates})
\begin{equation}
\label{final}
|\psi(-Z)\rangle=\sum_{n=0}^\infty (-1)^n\frac{d^n}{(2^nn!)^{1/2}}e^{-d^2/4}|n\rangle.
\end{equation}
According to (\ref{amplitude}), the corresponding intensity output at the $n$th waveguide is
\begin{equation}
\label{intensity}
|C_n(Z)|^2=\frac{d^{2n}}{2^nn!}e^{-d^2/2},
\end{equation}
the Poisson distribution corresponding to the coherent state (\ref{states}).
In order to develop some intuition, we plot in Fig. \ref{fig:outputs} the intensity output for $100$ waveguides ($n=0,1,2,\ldots,99$) designed to produce the desired output with the methods explained in the next paragraphs. The blue-squares curve corresponds to a displacement $d=5$, while the red-circles curve to $d=11$. Observe that the two outputs occupy different subsets of waveguides, and this is exactly the motivation for the design of such photonic lattices. We have verified that these output intensities are very close to the ideal values (\ref{intensity}) of a semi-infinite lattice, and the agreement is good as long as the light is concentrated away from the last (boundary) waveguide.

In order to design the modulating functions $a_i(z), i=1,2,3$, which produce the desired output, we will use the theory of Lewis-Riesenfeld invariants \cite{Lewis69}. Specifically, it is well known \cite{Leach77,TorronteguiPRA11,Schaff11} and can be directly verified that for the evolution (\ref{evolution}) under the Hamiltonian (\ref{Oscillator}) with $m(t)=1$, the following operator
\begin{equation}
\label{invariant}
\hat{I}=\frac{1}{2}[b(\hat{p}-\dot{a})-\dot{b}(\hat{q}-a)]^2+\frac{1}{2}\left(\frac{\hat{q}-a}{b}\right)^2
\end{equation}
with $a(t),b(t)$ satisfying the differential equations
\begin{eqnarray}
\label{a}\ddot{a}+\omega^2(t)a&=&f(t),\\
\label{b}\ddot{b}+\omega^2(t)b&=&\frac{1}{b^3},
\end{eqnarray}
is an invariant, i.e.
\begin{equation}
d_t\hat{I}=\partial_t\hat{I}+i[\hat{H},\hat{I}]=0.
\end{equation}
Note that the relation between the specific photonic lattice and Lewis-Riesenfeld invariants has been pointed out in \cite{Rodriguez14}, but here we use an invariant different than the one used there. The time-dependent eigenfunctions of $\hat{I}$ are \cite{TorronteguiPRA11,Schaff11}
\begin{equation}
\label{eigenfunctions}
|\lambda_n\rangle=e^{i[\dot{b}q^2/2b+(\dot{a}b-a\dot{b})q/b]}\frac{1}{b^{1/2}}\phi_n\left(\frac{q-a}{b}\right)
\end{equation}
where $\phi_n(q)=|n\rangle$ are given in (\ref{eigenstates}). They satisfy
\begin{equation}
\label{eigenvalues}
\hat{I}|\lambda_n\rangle=\lambda_n|\lambda_n\rangle,\qquad\lambda_n=n+\frac{1}{2},
\end{equation}
where the eigenvalues $\lambda_n$ are time-independent because $\hat{I}$ is invariant and real since $\hat{I}$ is Hermitian. It is clear from (\ref{eigenfunctions}) that (\ref{a}), (\ref{b}) are the equations for the mean position and the corresponding fluctuations, respectively. They describe the controllable part of the dynamics \cite{Mirrahimi04,Bloch10} which could have been obtained using Ehrenfest equations \cite{Choi13}. Note that $e^{i\theta_n(t)}|\lambda_n\rangle$ are also eigenfunctions of $\hat{I}$, and if the phases $\theta_n(t)$ are chosen to satisfy $d\theta_n/dt=\langle\lambda_n|i\partial_t-\hat{H}|\lambda_n\rangle$ or, specifically
\begin{equation}
\label{phases}
\theta_n(t)=-\int_0^t\left[\frac{\lambda_n}{b^2}+\frac{(\dot{a}b-a\dot{b})^2}{2b^2}-\frac{a^2}{2b^4}\right]d\tau,
\end{equation}
then the general solution of (\ref{evolution}) can be expressed as
\begin{equation}
\label{solution}
|\psi(t)\rangle=\sum_{n=0}^\infty c_ne^{i\theta_n(t)}|\lambda_n\rangle,
\end{equation}
where $c_n$ are constant coefficients \cite{TorronteguiPRA11,Schaff11}.

\begin{figure*}[t!]
 \centering
		\begin{tabular}{cc}
     	\subfigure[$\ $Displacement $d=5$, spread $\sigma=1$]{
	            \label{fig:transport1}
	            \includegraphics[width=.35\linewidth]{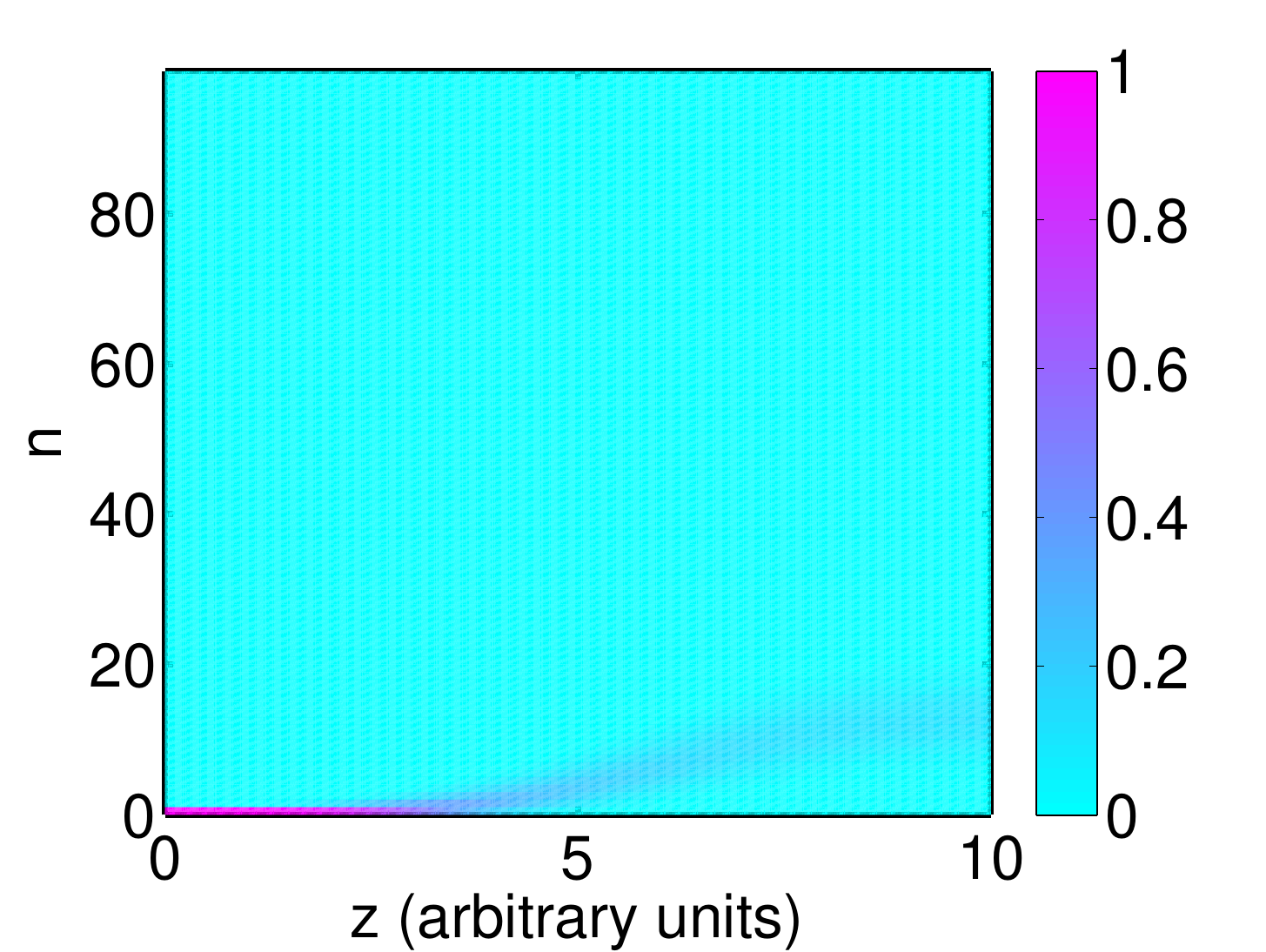}} &
       \subfigure[$\ $Displacement $d=5$, spread $\sigma=0.6$]{
	            \label{fig:transport2}
	            \includegraphics[width=.35\linewidth]{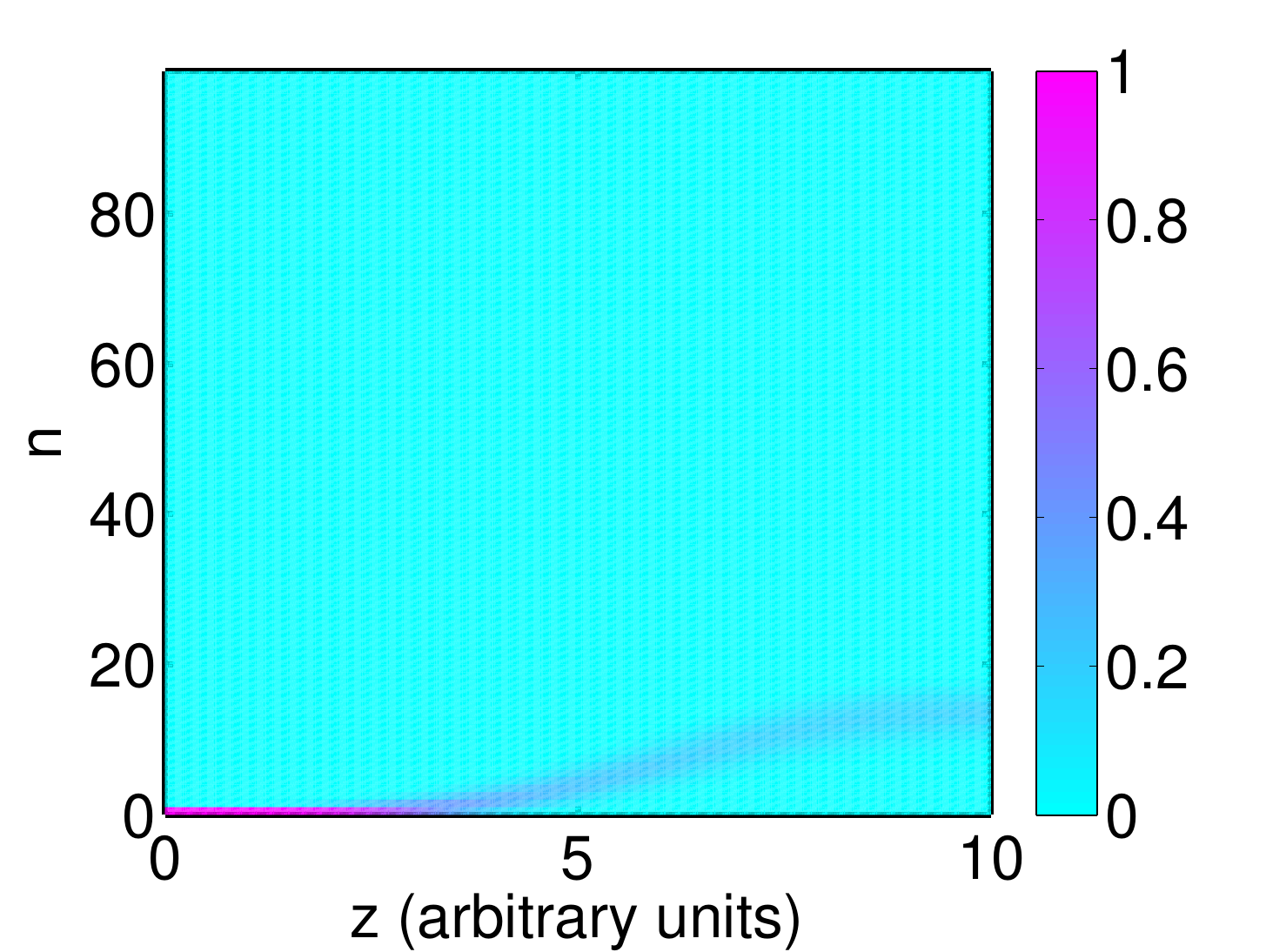}}\\
        \subfigure[$\ $Displacement $d=11$, spread $\sigma=1$]{
	            \label{fig:transport3}
	            \includegraphics[width=.35\linewidth]{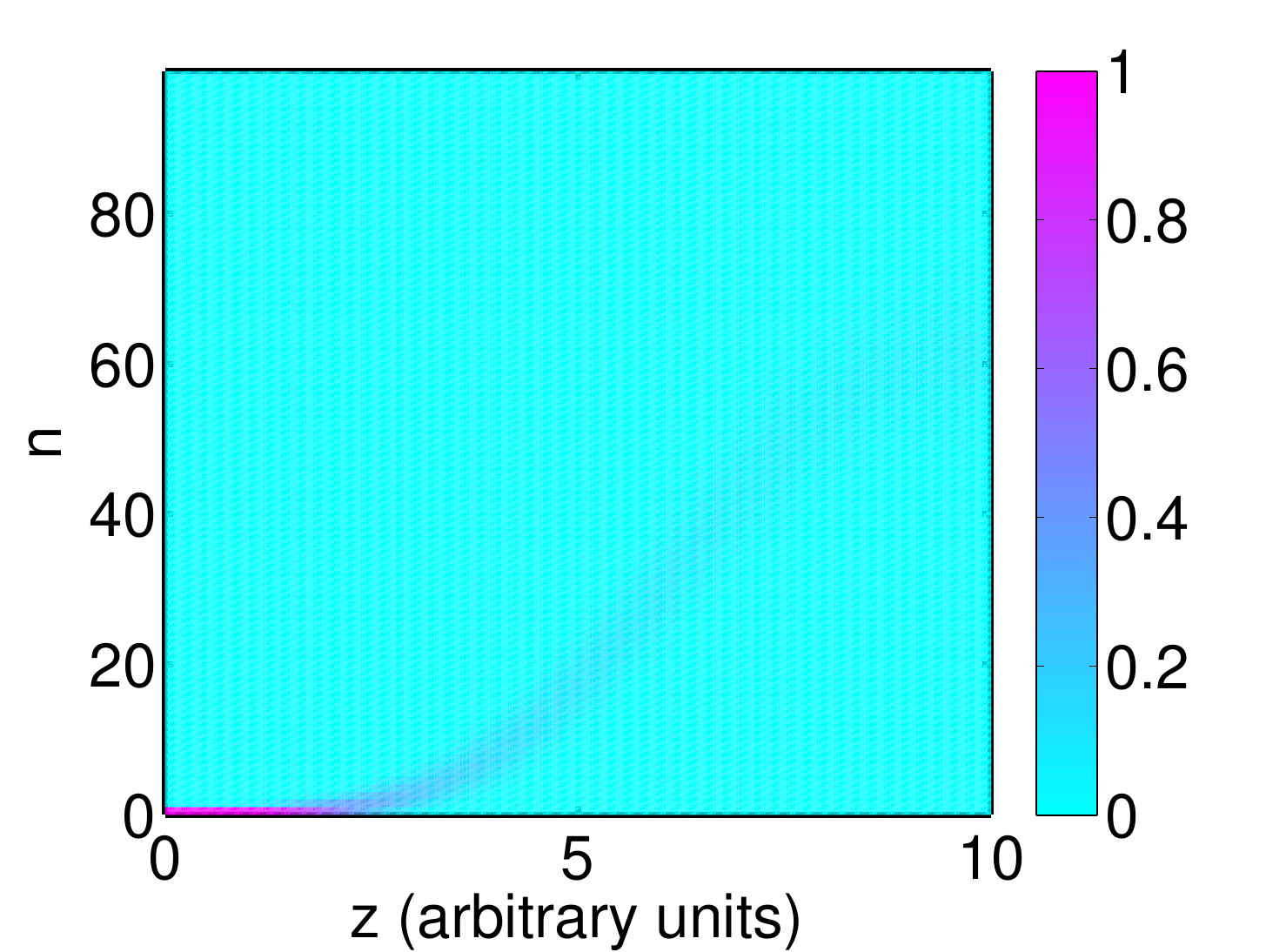}} &
        \subfigure[$\ $Displacement $d=11$, spread $\sigma=0.6$]{
	            \label{fig:transport4}
	            \includegraphics[width=.35\linewidth]{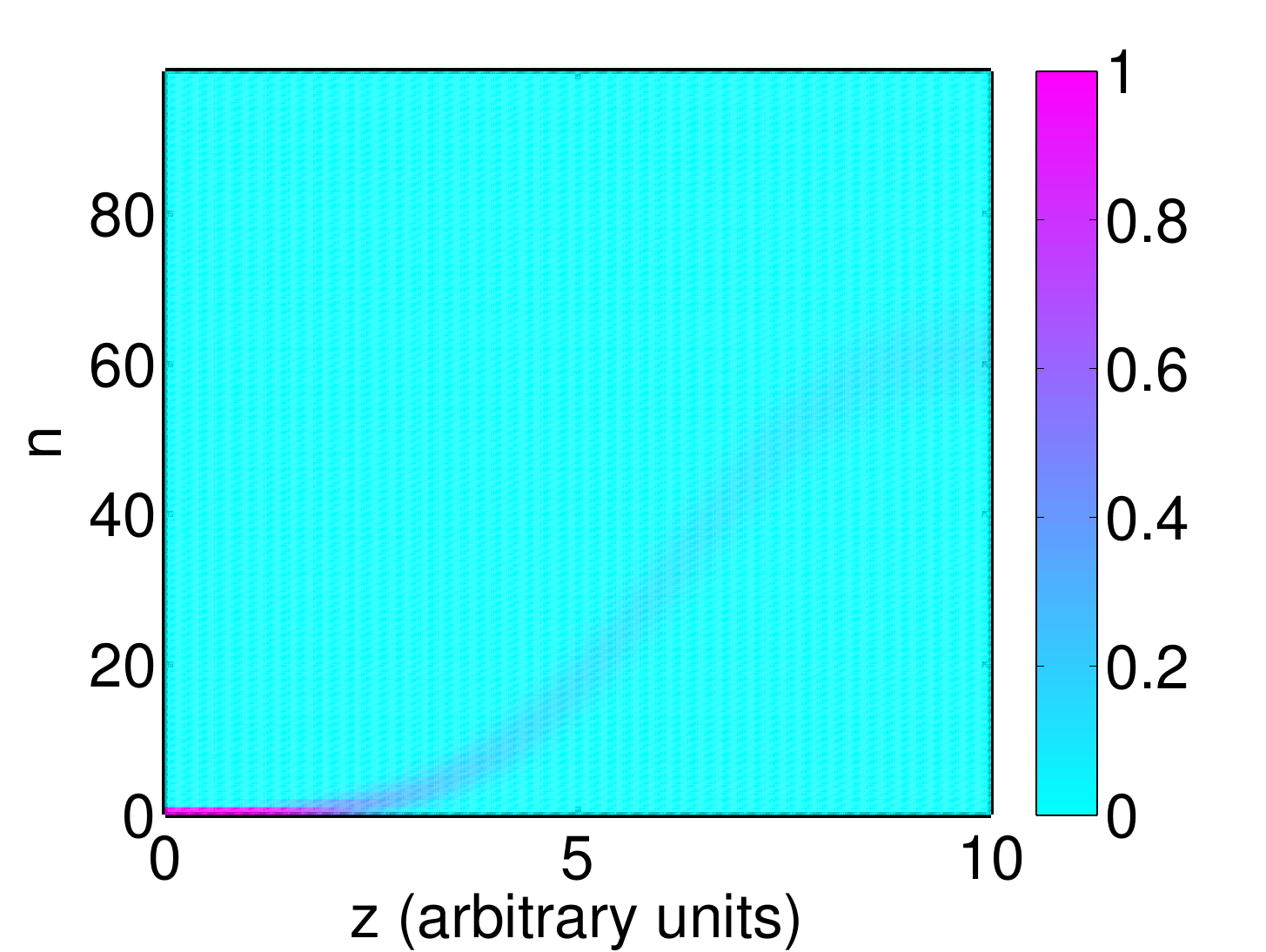}}\\
        \subfigure[$\ $Modulating functions for $\sigma=1$]{
	            \label{fig:transport5}
	            \includegraphics[width=.35\linewidth]{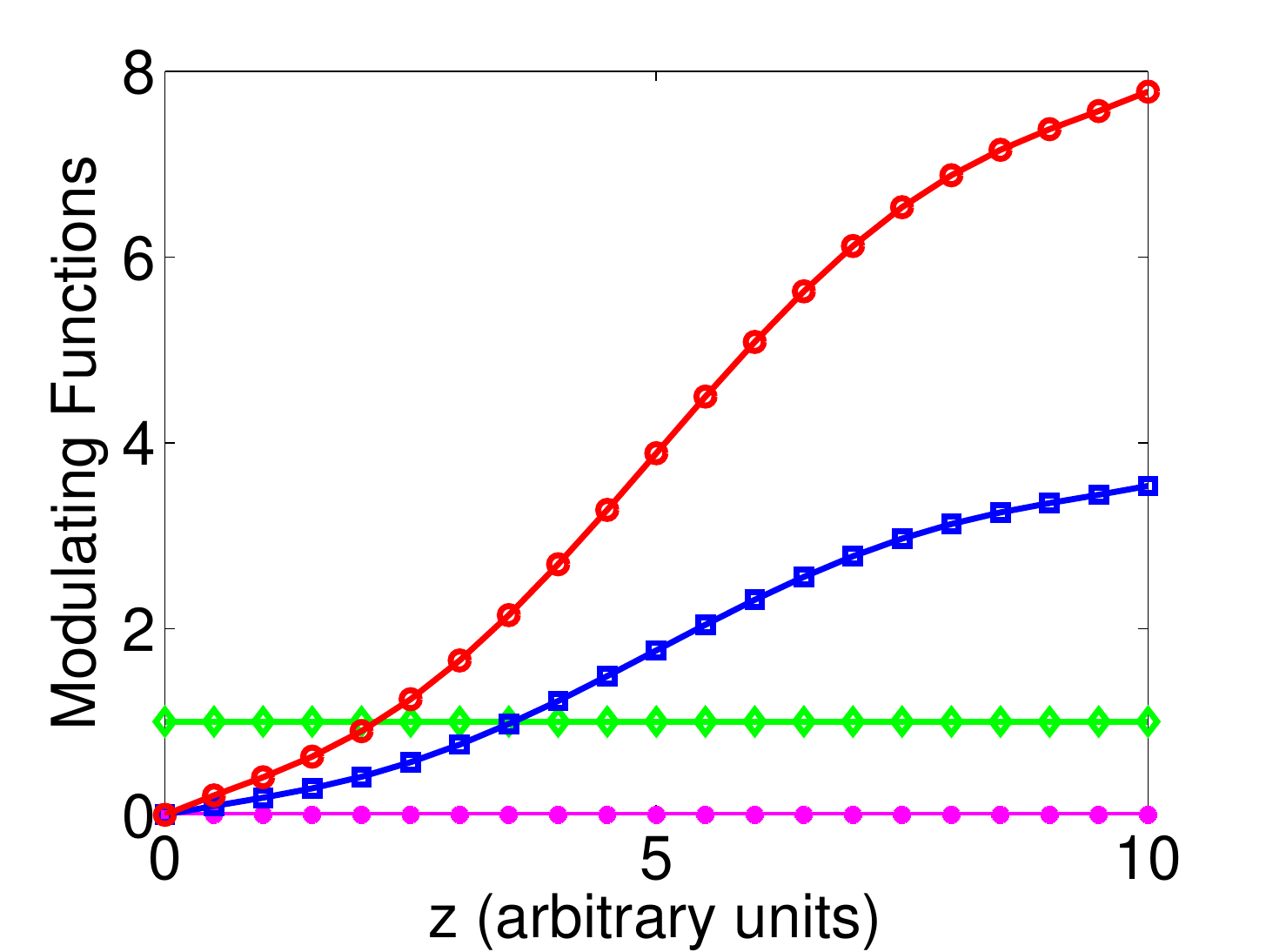}} &
        \subfigure[$\ $Modulating functions for $\sigma=0.6$]{
	            \label{fig:transport6}
	            \includegraphics[width=.35\linewidth]{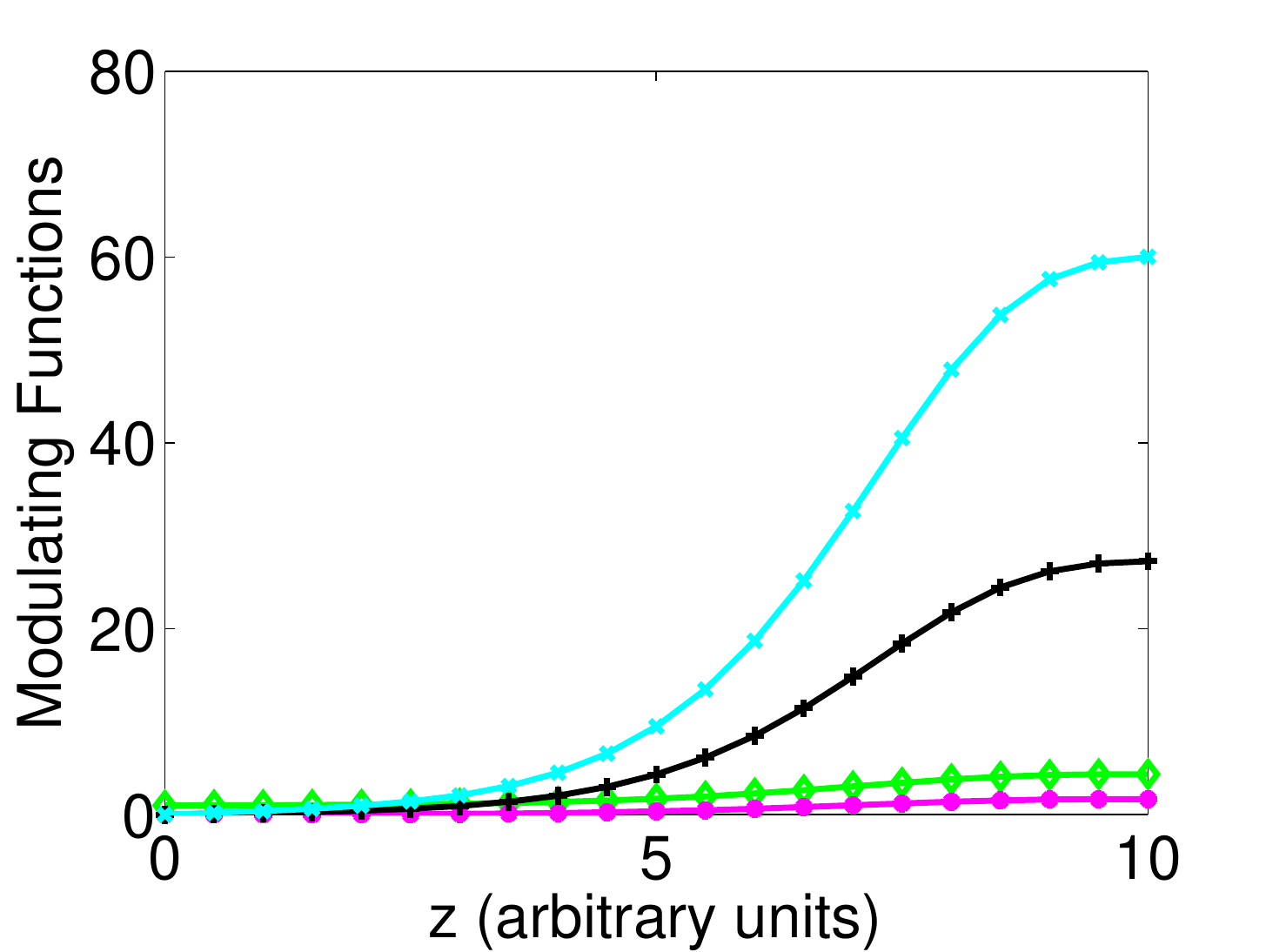}}
		\end{tabular}
\caption{(Color online) (a, b, c, d) Intensity propagation for a distance $Z=10$ when light is injected in the $0$th waveguide of a photonic lattice composed by $100$ waveguides. For (a, c) the lattice is designed to produce the target final state (\ref{states}) with $d=5,11$, respectively, while for (b, d) the final target state is (\ref{enhanced_final}) with common $\sigma=0.6$ and $d=5,11$. Observe that the spread of the output is considerably reduced in the right column. (e, f) Lattice parameters producing the propagation patterns of the corresponding columns. For each column, $a_0(z)$ (green diamonds) and $a_2(z)$ (magenta stars)  are common, while $a_1(z)$ is different and is plotted with colours and marks in symphony with those used in Fig. \ref{fig:outputs} (blue squares for $d=5,\sigma=1$, red circles for $d=11,\sigma=1$, black crosses for $d=5,\sigma=0.6$, cyan x-marks for $d=11,\sigma=0.6$).}
\label{fig:transport}
\end{figure*}

We now show how we can use Eq. (\ref{solution}) in the framework of shortcuts to adiabaticity \cite{shortcuts13} to design the photonic lattice. Recall that the essence of the method is to find a path connecting the initial and final eigenstates of a time-dependent Hamiltonian without following the instantaneous eigenstates, producing an effectively adiabatic evolution.
If we set $c_0=1, c_n=0$ for $n>0$, and the boundary conditions
\begin{eqnarray}
\label{boundary_a}a(0)=0,\quad&a(T)=-d, \quad&\dot{a}(0)=\dot{a}(T)=0,\\
\label{boundary_b}b(0)=1,\quad&b(T)=1, \quad&\dot{b}(0)=\dot{b}(T)=0,
\end{eqnarray}
where $T=-Z$ corresponds to the final distance, then from Eqs. (\ref{solution}), (\ref{eigenfunctions}) and (\ref{eigenstates}) we find that the initial and final states coincide with those in (\ref{states}), except an unimportant phase factor $e^{i\theta_0(T)}$ for the final state. We additionally require that the initial and final states are eigenstates of the instantaneous Hamiltonian at $t=0,T$. This requirement can be met when $[I(0),H(0)]=[I(T),H(T)]=0$ (recall that the initial and final states are already eigenstates of $I$ at $t=0,T$), and from these we obtain the additional boundary conditions $\omega(0)=\omega(T)=1, f(0)=0, f(T)=-d$. Using (\ref{a}), (\ref{boundary_a}) and (\ref{b}), (\ref{boundary_b}), it is not hard to see that these additional conditions are equivalent to
\begin{eqnarray}
\label{addot}\ddot{a}(0)=\ddot{a}(T)&=&0,\\
\label{bddot}\ddot{b}(0)=\ddot{b}(T)&=&0.
\end{eqnarray}
We next find $a(t),b(t)$ satisfying the boundary conditions (\ref{boundary_a}), (\ref{addot}) and (\ref{boundary_b}), (\ref{bddot}), respectively. Observe that we can simply choose
$b(t)=1$, while a polynomial interpolating function for $a$ is \cite{TorronteguiPRA11}
\begin{equation}
\label{poly}a(t)=-d(6s^5-15s^4+10s^3),
\end{equation}
where $s=t/T$. %Note that the above methodology, to connect the initial and final eigenstates of a time-dependent Hamiltonian without following the instantaneous eigenstates, producing an effectively adiabatic evolution, is called \emph{shortcut to adiabaticity} and has found several applications \cite{shortcuts13}. The paths obtained by this method are in general quite robust \cite{Choi12}.

From Eq. (\ref{b}) with $b(t)=1$ we have $\omega(t)=1$ while recall that $m(t)=1$, thus $M(z)=\Omega(z)=1$ and from (\ref{M}), (\ref{W2}) we find $a_0(z)=1, a_2(z)=0$.
From Eq. (\ref{a}) we find $f(t)=\ddot{a}+a$, which is a function of $s=t/T$, with $t=-z$ and $T=-Z$. It is not hard to see that $F(z)=f(-z)$ has the same functional dependence on $s=z/Z$. The modulating function $a_1(z)$ can be determined from Eq. (\ref{F}). In Fig. \ref{fig:transport1} we simulate the intensity propagation of light injected in the $0$th waveguide for a distance $Z=10$ and for a displacement $d=5$, while in Fig. \ref{fig:transport3} we use $d=11$. Observe that the two outputs are localized in different blocks of waveguides. The modulating functions producing these results are shown in Fig. \ref{fig:transport5}. For both cases it is $a_0(z)=1$ (green diamonds) and $a_2(z)=0$ (magenta stars). The function $a_1(z)$ differentiates the two cases and is depicted with blue squares for $d=5$ and red circles for $d=11$, in symphony with the colours and marks used in Fig. \ref{fig:outputs}.

Observe that for both values of the displacement parameter $d$ the output spans several waveguides. We can reduce the spread of the output intensity if we exploit the modulating functions $a_0(z)$ and $a_2(z)$, which were taken constant before, to control $b(t)$. Specifically, if we require the final value $b(T)=\sigma<1$, keeping the rest of the boundary conditions (\ref{boundary_b}), (\ref{bddot}) the same, then the final wavefunction becomes
\begin{equation}
\label{enhanced_final}|\psi(-Z)\rangle=\frac{1}{\sigma^{1/2}\pi^{1/4}}e^{-(q+d)^2/2\sigma^2}
\end{equation}
which is a sharper Gaussian than (\ref{states}) for $\sigma<1$. The following function satisfies the modified boundary conditions for $b$ \cite{Chen10}
\begin{equation}
\label{bpoly}b(t)=(\sigma-1)(6s^5-15s^4+10s^3)+1,
\end{equation}
where again $s=t/T$, while we use the previous polynomial (\ref{poly}) for $a(t)$. From (\ref{b}) we find $\omega^2(t)=(1/b^3-\ddot{b})/b$ and use it in (\ref{a}) to obtain $f(t)=\ddot{a}+\omega^2a$, while we remind that $m(t)=1$. As before, $\Omega(z),F(z)$ have the same form with $\omega(t),f(t)$ as functions of $s=t/T=z/Z$. In Fig. \ref{fig:outputs} we plot the intensity output for the same propagation distance $Z=10$ and displacement values $d=5$ (black crosses) and $d=11$ (cyan x-marks) as before, while we take $\sigma=0.6$ for both cases. In Figs. \ref{fig:transport2}, \ref{fig:transport4} we plot the light intensity propagation for these two cases, $d=5,\sigma=0.6$ and $d=11,\sigma=0.6$, respectively. Observe that the spread of the output has been considerably reduced compared to the Figs. \ref{fig:transport1}, \ref{fig:transport3}. In Fig. \ref{fig:transport6} we plot the corresponding modulating functions. Observe that $a_0(z)$ (green diamonds) and $a_2(z)$ (magenta stars) are common since $\sigma$ is common, while $a_1(z)$ differentiates the two cases and is depicted with black crosses for $d=5$ and cyan x-marks for $d=11$, in symphony with the colours and marks used in Fig. \ref{fig:outputs}.

We conclude this section with a short discussion about the size of the waveguide length $Z$. In theory it can become arbitrarily small \cite{Chen10}, but in practice there are always experimental limitations, for example on the size of the lattice characteristics, which impose a minimum necessary value for $Z$ \cite{Stefanatos10}. In order to maintain the robustness properties of shortcuts to adiabaticity, it is better to avoid waveguide lengths close to this minimum \cite{Choi12}. Yet, the necessary length is smaller than that required by devices based on purely adiabatic schemes \cite{Tseng12,Chien13,Tseng13,Martinez14}.

%%%%%%%%%%%%%%%%%%%%%%%%%%%%%%%%%%%%%%%%%%%%%%%%%%%%%%%%%%%%%%%%%%%%%%%%%%%%%%%%
\section{CONCLUSION AND OUTLOOK}

\label{sec:conclusion}

In this paper we have used shortcuts to adiabaticity to design a photonic lattice which can direct the input light to different locations at the output, depending on the choice of the functions modulating the lattice characteristics (index of refraction, first and second neighbor couplings). If these functions can be altered in time, in a physical context different than the evanescently coupled waveguides, then this lattice structure can also implement a switch. The proposed device is expected to be robust because of the adiabatic nature of the design method and also to have a smaller footprint than purely adiabatic devices. A very interesting extension of the present work is to use optimal control \cite{Pontryagin} for the dynamical system describing the lattice, in order to find the modulating functions (control inputs) which produce more sophisticated output patterns. Specifically, the system described by Eqs. (\ref{model}) with $n$ finite, $a_0(z)$ constant and $a_2(z)=0$, which is a truncation of a quantum harmonic oscillator with dipole interaction, has been proven to be controllable \cite{Schirmer01}. This means that any initial state at $z=0$ can evolve to any final state at $z=Z$, with the appropriate choice of the control $a_1(z)$. Depending on the initial and final states the connecting control can be quite complicated, and one has to use numerical methods in order to calculate it. For a bilinear control system like (\ref{model}) (linear in both the states $C_n$ and the control $a_1$), an efficient numerical method is described in \cite{Aganovic94}. Note that if $a_0(z)$ and $a_2(z)$ are also allowed to vary with $z$, instead of being fixed, they can serve as extra control inputs (degrees of freedom) which can be further exploited. The suggested methodology provides a concrete path for the design of lattices with complex input-output characteristics, as well as for the engineering of classical analogues of quantum states. The present paper is a nice prelude to this very promising future work.

\end{document}